\documentclass[prd,twocolumn,showpacs,preprintnumbers,aps,nofootinbib]{revtex4-1}

\usepackage{graphicx}
\usepackage{dcolumn}
\usepackage{bm}
\usepackage{amsmath,amssymb,amsfonts}
\usepackage{latexsym}

\usepackage{color}

\def\nn    {\nonumber}

\begin{document}


\title{\boldmath
The Coming Decade of $h \to \tau\mu$ and $\tau \to \mu\gamma$ Interplay
 in $\tau$ Flavor Violation Search
}

\author{Wei-Shu Hou and Girish Kumar}
\affiliation{
Department of Physics, National Taiwan University, Taipei 10617, Taiwan}
\bigskip


\begin{abstract} 
We explore the interplay between $h(125) \to \tau\mu$ search at the LHC
and $\tau \to \mu\gamma$ at the up and coming Belle~II experiment,
in context of the general two Higgs doublet model with
extra Yukawa couplings such as $\rho_{\tau\mu}$.
The search for $h \to \tau\mu$ constrains $\rho_{\tau\mu} \cos\gamma$,
where $\cos\gamma$ is the $h$--$H$ mixing angle of $h$ 
with the exotic $CP$-even scalar $H$.
For $\tau \to \mu\gamma$, we define the ``BSM-benchmark'' by setting 
the extra top Yukawa coupling $\rho_{tt} = 1\ (\cong \lambda_t)$ in two-loop diagrams, 
and $\cos\gamma = 0$ to decouple $h$.
We show that this leading effect due to $H$ and $CP$-odd scalar $A$
can be readily probed by Belle~II,
even for the conservative value of $\rho_{\tau\mu} = 0.7\lambda_\tau$.
We define the subleading ``$h$-benchmark'' 
by setting $\rho_{tt} = 0$ in two-loop diagrams, 
and take the conservative maximal value of $\cos\gamma = 0.2$. 
We show that it falls beyond Belle~II reach, but can 
interfere with the\ BSM-benchmark effect, which in principle probes the phase of $\rho_{tt}$.
We further show that the one-loop $H$, $A$ effect,
proportional to $\rho_{\tau\mu}\rho_{\tau\tau}$ in amplitude,
is beyond the sensitivity of Belle~II to probe,
even for $\rho_{\tau\tau}$ as large as $3\lambda_\tau$.
With the working assumption that $\rho_{32}^f, \rho_{33}^f  = {\cal O}(\lambda_3^f)$
for all charged fermions $f$, we find good discovery potential for 
both $\tau$ lepton flavor violation searches in the coming decade.
\end{abstract}

\maketitle


\section{Introduction}

Ever since the discovery of the muon and finding an empirical
``muon number'' that is separate from the electron number, 
the issue of lepton number violation has been pursued. 
Extending to the third generation of leptons, 
the B factory era closed with the bound~\cite{PDG},
\begin{align}
{\cal B}(\tau \to \mu\gamma) < 4.4 \times 10^{-8}, \quad\quad ({\rm PDG18})
\label{taumugam_PDG}
\end{align}
which is from the BaBar experiment~\cite{Aubert:2009ag}
and based on $\sim 0.96 \times 10^9$ $\tau$ decays. 
The Belle experiment has an earlier result~\cite{Hayasaka:2007vc}
 at $4.5 \times 10^{-8}$, based on $\sim 0.48 \times 10^9$ $\tau$ decays,
but somehow has not updated.
The Belle~II experiment, which has commenced B physics running,
aims at improving the bound 
by a factor of 100,
which we take conservatively as $10^{-9}$~\cite{Kou:2018nap}.
Thus, there is potential for discovery in the coming decade.

\begin{figure*}[t]
\center
\includegraphics[width=0.25 \textwidth]{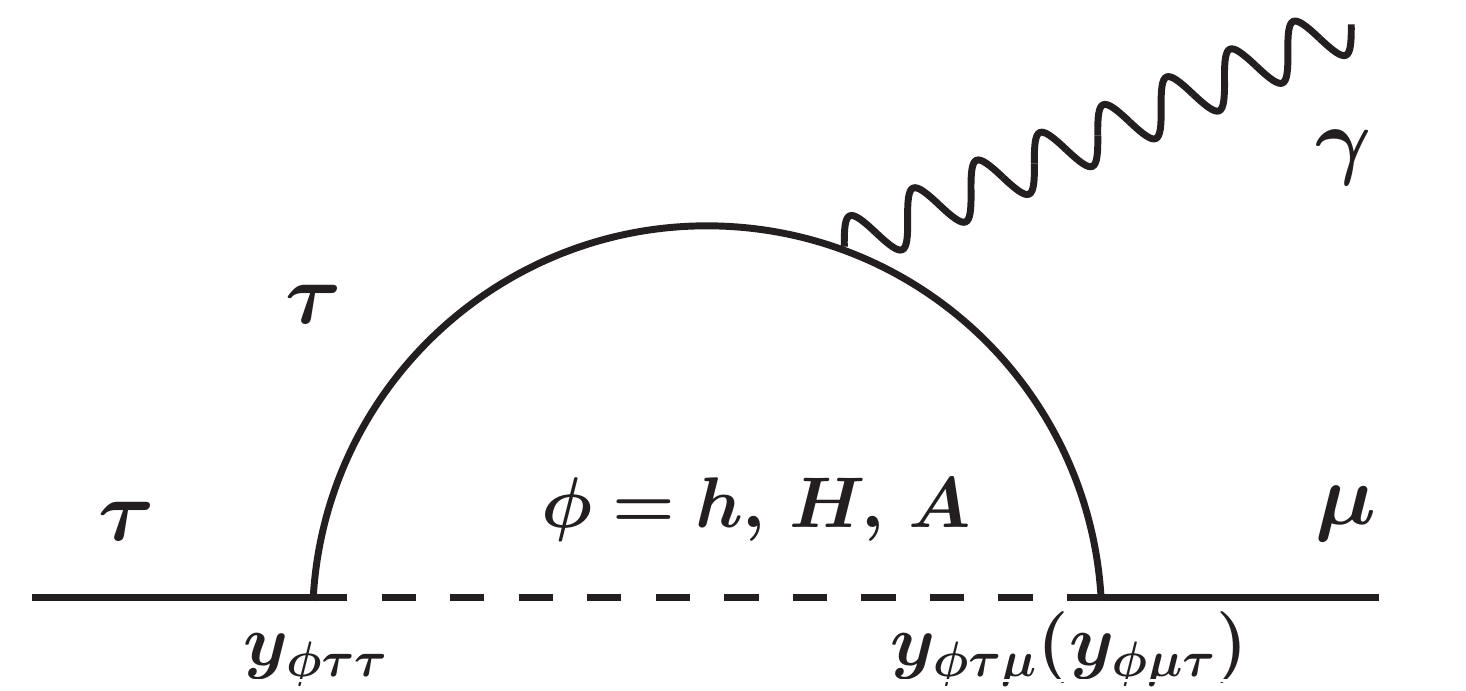}
\hskip0.35cm
\includegraphics[width=0.25 \textwidth]{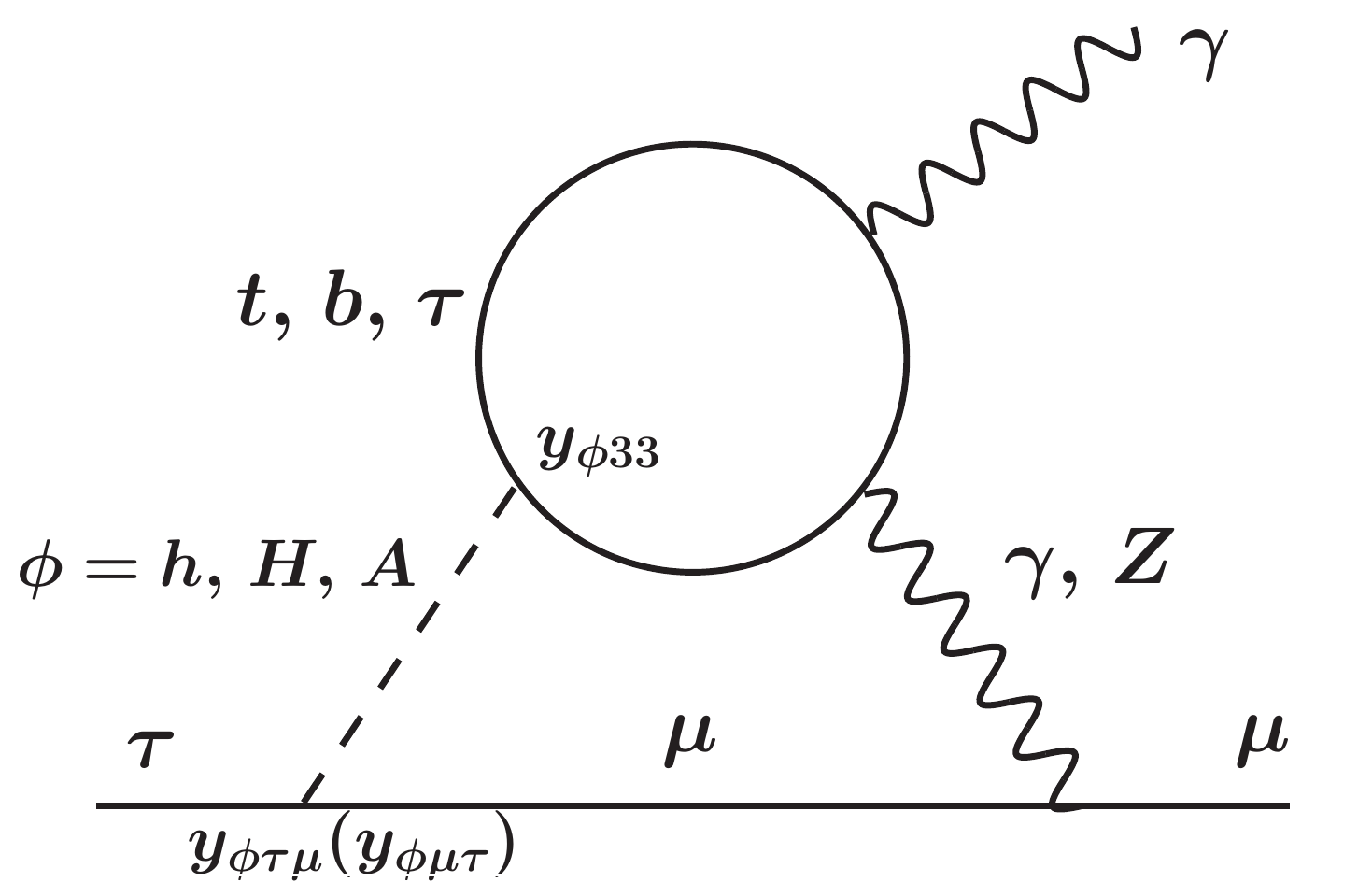}
\hskip0.35cm
\includegraphics[width=0.25 \textwidth]{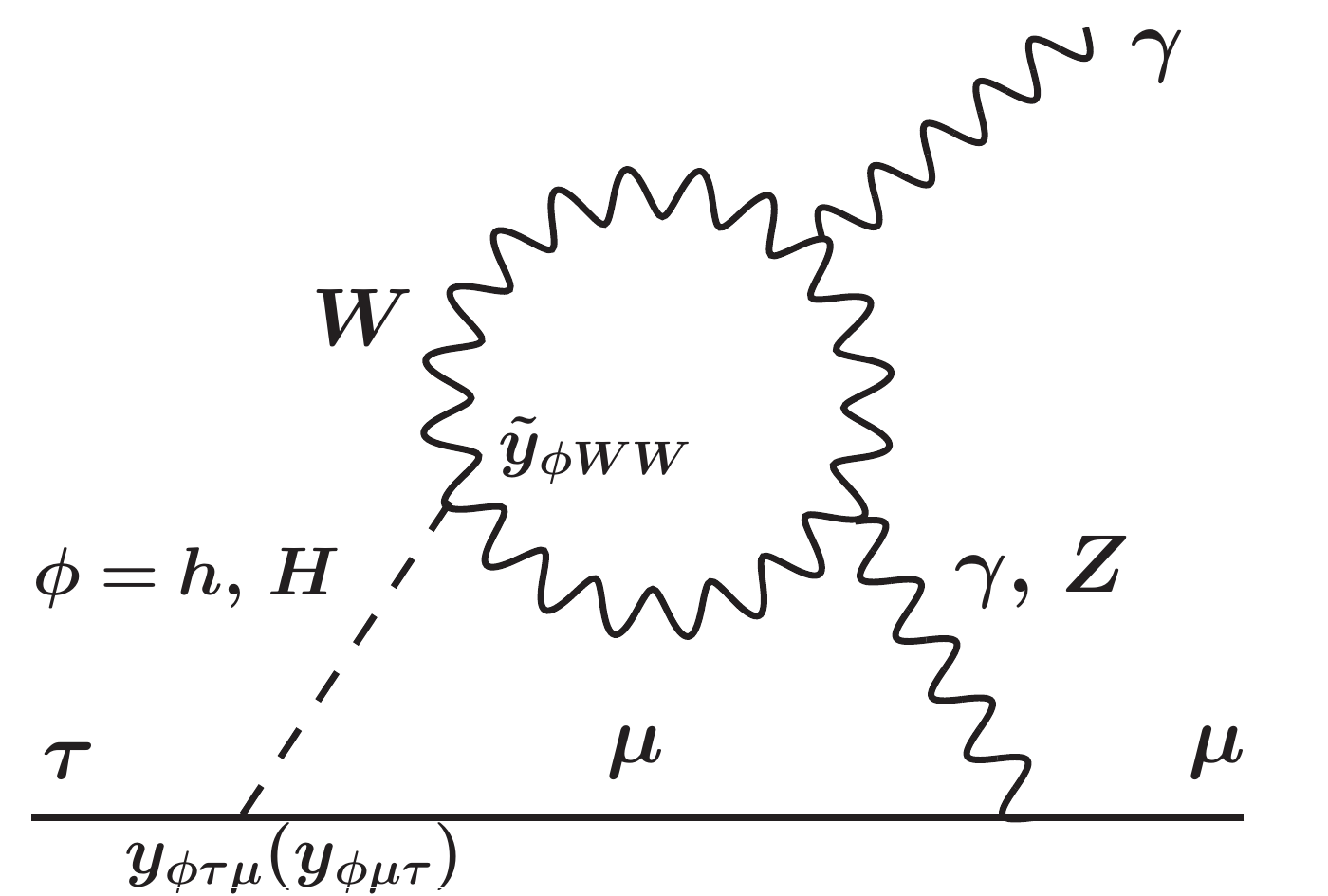}
\caption{
One-loop, two-loop fermion and two-loop $W$ diagrams for $\tau\to \mu\gamma$.}
\label{feyndiag}
\end{figure*}

The discovery of the 125 GeV scalar boson $h$~\cite{h125_discovery}
completes the last piece of the Standard Model (SM),
and is a triumph of the Large Hadron Collider (LHC). 
With LHC Run 1 data at 8 TeV collision energy,
the CMS experiment found~\cite{Khachatryan:2015kon} 
an intriguing 2$\sigma$ hint for the $\tau$ lepton flavor violating
 ($\tau$LFV) $h \to  \tau\mu$ process, 
which subsequently disappeared~\cite{Sirunyan:2017xzt} 
with 13 TeV data at Run~2,
\begin{align}
{\cal B}(h \to  \tau\mu) < 0.25 \%. \quad\quad ({\rm CMS18})
\label{h-taumugamCMS}
\end{align}
Recently, with similar amount of data at $\sim 36$ fb$^{-1}$, 
the ATLAS experiment reported~\cite{Aad:2019ugc} a consistent bound of 0.28\%. 
As this is still less than 1/3 of the full Run~2 data at hand
 for each experiment, updates are expected.
Furthermore, the scheduled Run~3 for 2021--2024 would
likely add twice more data than Run~2.
Thus, scaling naively by statistics, and 
assuming that ATLAS and CMS would make a combined analysis
before the start of High Luminosity LHC (HL-LHC) 
targeted for 2028 --- especially if there is some hint! --- 
the limit could reach 0.05\%, with corresponding 
discovery potential.

Thus, there is much to look forward to in the coming decade
on the $\tau$LFV front.
This paper aims at elucidating the relevant contributions
and parameters of importance, enhancing what has been discussed already.

To have $\tau\mu h$ couplings, the framework is 
a two Higgs doublet model (2HDM)~\cite{Branco:2011iw}
without a $Z_2$ symmetry to forbid flavor changing neutral Higgs (FCNH) couplings,
which was dubbed ``Model III''~\cite{Hou:1991un} 
(distinct from Models I \& II under $Z_2$ symmetry) of 2HDM a long time ago.
There is a vast amount of theory work on $\tau$LFV that 
we cannot possibly do justice to, and we refer to the recent 
mini-reviw of Vicente~\cite{Vicente:2019ykr}.
Instead, let us trace some major steps in the phenomenological development.

The template for discussing $\tau \to \mu\gamma$ decay can be traced to  
the work of Chang, Hou and Keung~\cite{Chang:1993kw},
which studied the $\mu \to e\gamma$ transition in the context of 2HDM~III.
The paper stressed that the top contribution to the two-loop 
Bjorken-Weinberg (or Barr-Zee) mechanism, 
by bringing in the intrinsically larger extra top Yukawa coupling, 
can be much larger than the one-loop effect
 (middle and left diagrams of Fig.~\ref{feyndiag}).
One just changes the formulas from $\mu \to e$ labels to $\tau\to \mu$,
which was followed by all subsequent workers.

The $h \to \tau\mu$ process was proposed by Han and Marfatia~\cite{Han:2000jz}
at the start of Tevatron Run II, also in the context of 2HDM III.
As the Tevatron era was coming to an end, and at the dawn of the LHC,
Davidson and Grenier~\cite{Davidson:2010xv} took interest in 
$h \to \tau\mu$ at colliders, and emphasized the link with 
$\tau \to \mu\gamma$ bound from B factories as an important constraint.
The work, however, was oriented towards the lepton perspective.
Extending from earlier and more general work~\cite{Davidson:2005cw}, 
the authors defined $\tan\beta_\tau = \rho_{\tau\tau}/\lambda_\tau$,
 where $\rho_{\tau\tau}$ is the extra diagonal $\tau$ Yukawa coupling,
 and $\lambda_\tau = \sqrt2 m_\tau/v$ ($v \cong 246$ GeV) is the
 $\tau$ Yukawa coupling of SM,
and used $\tan\beta_\tau$ in place of the familiar $\tan\beta$
of 2HDM with $Z_2$ (e.g. the well known 2HDM II).
Knowing that, without a $Z_2$ symmetry, $\tan\beta$ as the
ratio of v.e.v.'s of the two Higgs doublets is not a physical parameter,
the authors sought substitute in language and usage, 
but it should be clear that the ratio of Yukawa couplings is quite a different thing.
The authors further extended $\tan\beta_\tau$ into the quark sector, 
which is a strong assumption. 
Adopting this, the early work of Aristizabal Sierra and Vicente~\cite{Sierra:2014nqa}
in addressing the CMS hint of $h \to \tau\mu$ excess~\cite{Khachatryan:2015kon} 
allowed $\tan\beta_\tau = \rho_{\tau\tau}/\lambda_\tau$
to be as large as 40, 
i.e. the extra $\tau$ Yukawa coupling could be
almost half the strength of the top Yukawa coupling.
We will not take this lepton-biased view, and let
extra top Yukawa couplings be independent parameters.

The CMS study that showed excess~\cite{Khachatryan:2015kon} was in fact 
inspired by the work of Harnik, Kopp and  Zupan~\cite{Harnik:2012pb}. 
While using the formulas of Ref.~\cite{Chang:1993kw} as usual
to study the $\tau \to \mu\gamma$ constraint on the $\tau\mu h$ coupling,
they showed that a direct search for $h \to \tau\mu$ at the LHC 
would quickly become more sensitive.
The paper, however, used the language of Cheng and Sher~\cite{Cheng:1987rs},
which was adopted also in the CMS papers.
While capturing the mass-mixing hierarchy suppression (Model III~\cite{Hou:1991un}) 
of FCNH for low energy processes, the Cheng-Sher ansatz missed one element,
that the FCNH couplings are associated with the exotic (non-mass-giving) Higgs doublet, 
and would enter the coupling of the SM-like $h$ to e.g. $\tau\mu$
by the $h$--$H$ mixing angle between the two $CP$-even scalars.
Thus, the $\tau\mu h$ coupling reads as $\rho_{\tau\mu}\cos(\beta-\alpha)$,
where for the time being we retain the familiar notation of 2HDM~II.

The latter approach was adopted by Omura, Senaha and Tobe~\cite{Omura:2015xcg} 
in correlating $h \to \tau\mu$ excess with predictions for $\tau\to \mu\gamma$,
where they entertained $\rho_{\tau\mu}$, $\rho_{\tau\tau}$ up to $10\lambda_\tau$
for $c_{\beta-\alpha} \equiv \cos(\beta-\alpha) \simeq 0.1$.
The point is, when the CMS excess disappeared with more data,
it could just be due to the smallness of $c_{\beta-\alpha}$
(the phenomenon of {\it alignment}),
rather than demanding $\rho_{\tau\mu}$ to be small.
Turning this around, the proposed search~\cite{Hou:2019grj} for
$H,\, A \to \tau\mu$ (where $A$ is the pseudoscalar) 
is not suppressed by alignment, or small $c_{\beta-\alpha}$.
The process has now already been searched for by CMS~\cite{Sirunyan:2019shc},
 setting bounds.

We have mentioned quite a few parameters in our retracing  of
the development of $h \to \tau\mu$ and $\tau\to \mu\gamma$ decay studies.
The main goal of this paper is to elucidate the relevant vs less relevant parameters,
as the coming decade unfolds for the search of these two important $\tau$LFV processes,
to clarify the landscape.
Another motivation arose from the recent $H, A \to \tau\mu$ study~\cite{Hou:2019grj},
where constraints on $\rho_{tt}$
 (extra top Yukawa coupling that enters $\tau \to \mu\gamma$ at two-loop) 
and $\rho_{\tau\mu}$ from e.g. $\tau \to \mu\gamma$ 
was extracted by assuming $\rho_{tt}$ to be real, ``for simplicity''.
While this is a common, prevailing assumption, but 
just a couple of years prior, and before the hint for $h\to \tau\mu$ evaporated, 
it was pointed out~\cite{Fuyuto:2017ewj} 
that the complexity of $\rho_{tt}$ {\it could drive} 
the Baryon Asymmetry of the Universe (BAU).
With such big issues at stake, this paper explores the
possible effect of $\varphi_{tt} = {\rm arg}\, \rho_{tt}$,
which has not been explored before.
We shall call 2HDM III, or 2HDM without $Z_2$ symmetry
and where extra Yukawa couplings are allowed, the general 2HDM (g2HDM).

\section{
Parameters and Formulas
 in the General 2HDM}

In this paper we will take the masses of the physical 
$CP$-even scalars $h$, $H$, $CP$-odd scalar $A$, and charged scalar $H^+$ 
as given, and would not be concerned with details of the Higgs potential, 
which can be found e.g. in Ref.~\cite{Hou:2017hiw}.
The Yukawa couplings 
are~\cite{Davidson:2005cw,Hou:2017hiw}
\begin{align}
\mathcal{L} = 
 - & \frac{1}{\sqrt{2}} \sum_{f = u, d, \ell} 
 \bar f_{i} \Big[\big(\lambda^f_i \delta_{ij} s_\gamma + \rho^f_{ij} c_\gamma\big) h \nn\\
 & + \big(\lambda^f_i \delta_{ij} c_\gamma - \rho^f_{ij} s_\gamma\big)H
    - i\,{\rm sgn}(Q_f) \rho^f_{ij} A\Big]  R\, f_{j} \nn\\
 & - \bar{u}_i\left[(V\rho^d)_{ij} R-(\rho^{u\dagger}V)_{ij} L\right]d_j H^+ \nn\\
 &- \bar{\nu}_i\rho^\ell_{ij} R \, \ell_j H^+
 +{h.c.},
\label{eff}
\end{align}
where $i$, $j$ are generation indices that are summed over, 
$L, R = (1\mp\gamma_5)/2$ are projection operators, 
and $V$ is the Cabibbo-Kobayashi-Maskawa matrix.
Due to the very near degeneracy of the neutrinos for our processes, 
the corresponding matrix in lepton sector is taken as unity.
The shorthand notation of $c_\gamma \equiv \cos\gamma$ 
(and $s_\gamma \equiv \sin\gamma$) is the $h$--$H$ mixing angle, 
which corresponds to the usual $\cos(\beta - \alpha)$ in 2HDM II nomenclature.
The emergent {\it alignment} phenomenon,
that $h$ so closely resembles the SM Higgs boson~\cite{Khachatryan:2016vau},
implies that $c_\gamma$ is rather small.
But we do not quite know its value, which is especially true in g2HDM,
where more parameters exist compared with 2HDM II.
In the alignment limit of $c_\gamma \to 0$,
the couplings of $h$, including to vector bosons, do approach SM.
But as shown in Ref.~\cite{Hou:2017hiw},
small $c_\gamma$ need not imply small Higgs quartic couplings.
Thus, the prerequisite~\cite{Fuyuto:2017ewj} of 
${\cal O}(1)$ Higgs quartics for sake of 
first order electroweak phase transition for generating BAU, can be sustained.

The off-diagonal coupling $\rho_{\tau\mu}$ (and $\rho_{\mu\tau}$) enters 
the $\tau\to \mu\gamma$ and $h\to \tau\mu$ processes of interest.
Note that the first FCNH parameter studied directly at the LHC is $\rho_{tc}$ 
via $t \to ch$ decay~\cite{PDG}, which was 
pointed out already in Ref.~\cite{Hou:1991un} 
and reemphasized~\cite{Chen:2013qta} shortly after the $h(125)$ discovery.
Whether $h\to \tau\mu$ or $t \to ch$, the SM-like $h$ boson
picks up the FCNH coupling via a factor of $c_\gamma$, or $h$-$H$ mixing.
From hindsight, as discussed in Ref.~\cite{Hou:2017hiw},
the alignment phenomenon that emerged with full Run~1 data
can account for the absence so far of $t\to ch$ and $h \to \tau\mu$,
without the need of overly suppressing extra FCNH Yukawa couplings
 $\rho_{tc}$ or $\rho_{\tau\mu}$.
But since
\begin{align}
{\cal B}(h \to \tau\mu) =
 \frac{m_h c_\gamma^2}{16\pi\Gamma_h}
 (|\rho_{\tau\mu}|^2 + |\rho_{\mu\tau}|^2),
\end{align}
the bound of Eq.~(\ref{h-taumugamCMS}) places the constraint of
\begin{align}
 |\rho_{\tau\mu}\, c_\gamma| \lesssim 0.0014 \simeq 0.14\lambda_\tau,
\label{B_h-taumu}
\end{align}
where we have taken $|\rho_{\mu\tau}| = |\rho_{\tau\mu}|$ to simplify.
The two chiral couplings do not interfere.

As elucidated by Davidson and Grenier~\cite{Davidson:2010xv}
(from the template of Ref.~\cite{Chang:1993kw} for $\mu \to e\gamma$),
there are three distinct types of diagrams contributing to $\tau\to \mu\gamma$:
the one-loop diagram that pairs the necessary FCNH $\rho_{\tau\mu}$ coupling 
with a diagonal $\tau$ Yukawa coupling, be it the $\lambda_\tau$ of SM,
or the extra $\rho_{\tau\tau}$;
the two-loop Bjorken-Weinberg/Barr-Zee type of diagrams with top Yukawa,
be it $\lambda_t$, or $\rho_{tt}$;
and the two-loop $W$ diagram. 
The three type of diagrams are illustrated in Fig.~\ref{feyndiag}.
The $H^+$ effect is unimportant.
In these diagrams, we have labeled the vertices with the compact notation of
Ref.~\cite{Omura:2015xcg} (similar to Davidson and Grenier),
$-y_{\phi ij}^f \bar f_iRf_j\phi$ (h.c. implied)
 for $f = u, d, \ell$ and $\phi = h, H, A$, where
$y_{\phi ij}^f$ can be read off from Eq.~(\ref{eff}).
One can now see the two-loop mechanism constitutes an insertion
of $\phi \to \gamma\gamma$ (we shall neglect the $Z$ contribution),
which is similar to the $gg$ fusion production of $\phi$,
hence connecting with the $h \to \tau\mu$ and $H, A \to \tau\mu$ searches.

%
%
%
The branching fraction for $\tau\to\mu\gamma$ can be written as
\begin{equation}
	\frac{{\cal B}(\tau\to\mu\gamma)}{{\cal B}(\tau\to\mu\nu\bar\nu)} = \frac{48\pi^3\alpha}{G_F^2}\left(|A_L|^2+|A_R|^2\right),
\end{equation}
where ${\cal B}(\tau\to\mu\nu\bar\nu)= 17.39\%$ \cite{PDG},
and the chiral amplitudes $A_L$ and $A_R$, which do not interfere,
contribute equally under our simplifying assumption of $\rho_{\tau\mu} = \rho_{\mu\tau}$. 
The $A_L$ amplitude corresponding to the three type of diagrams in Fig.~\ref{feyndiag}
are (three separate sums)
\begin{widetext}
	\begin{equation}\begin{aligned} 
	A_{L} \simeq 
&   \sum_{\phi=h, H, A }
       \frac{\hat y_{\phi \tau \mu}^{\ast} \hat y_{\phi \tau \tau}^{\ast }}{8 \pi^{2}v^2}
       x_{\tau\phi} \bigl(\log x_{\phi\tau}- {3}/{2}\bigr) 
     - \sum_{\phi=h, H, A }
       \frac{N_{C} Q_{f}^2 \alpha}{4 \pi^{3} v^2} {\hat y_{\phi\tau\mu}^{\ell \,\ast}}
       \Bigl[\operatorname{Re} (\hat y_{\phi tt}) F_{H}(x_{t\phi})
             - i\operatorname{Im}(\hat y_{\phi tt}) F_{A}(x_{t \phi})\Bigr] \\
& + \sum_{\phi=h, H}
        \frac{ \alpha\,\tilde g_{\phi W W}\, \hat y_{\phi\tau\mu}^{\ast}}{32\sqrt2 \,\pi^3 v^2}
        \Bigl\{12 F_{H}(x_{W\phi})+{23} F_{A}(x_{W \phi})+{3} G(x_{W_{\phi}})+2x_{\phi W}\bigl[F_{H}(x_{W\phi})-F_{A}(x_{W\phi})\bigr]\Bigr\}\,,
	\end{aligned}\end{equation}
\end{widetext}
where $\hat y_{\phi \tau j} = y_{\phi \tau j}/\lambda_\tau$ 
(and likewise $\hat y_{\phi tt} = y_{\phi tt}/\lambda_t$), 
$x_{a b} = m_a^2/m_b^2$, $N_c$ is the number of colors, 
$\tilde{g}_{h WW} = s_\gamma$, $\tilde{g}_{H WW} = c_\gamma$, 
and the loop functions $F_H$, $F_A$ and $G$ can be found in, e.g. Ref.~\cite{Chang:1993kw}. 
%
%
We include only the $\phi\gamma\gamma$ vertex contributions 
and neglect $\phi Z\gamma$ vertex terms, 
as these are suppressed by $(1-4 \sin^2\,\theta_W)$,  
which amounts to $\sim 10\%$ variation 
 in our results.
%
The $b$ and $\tau$ contributions in the second sum
are suppressed by loop functions, as $x_{b\phi}$ and $x_{\tau\phi}$ are rather small.

\begin{figure*}[t]
\center
\includegraphics[width=0.425 \textwidth]{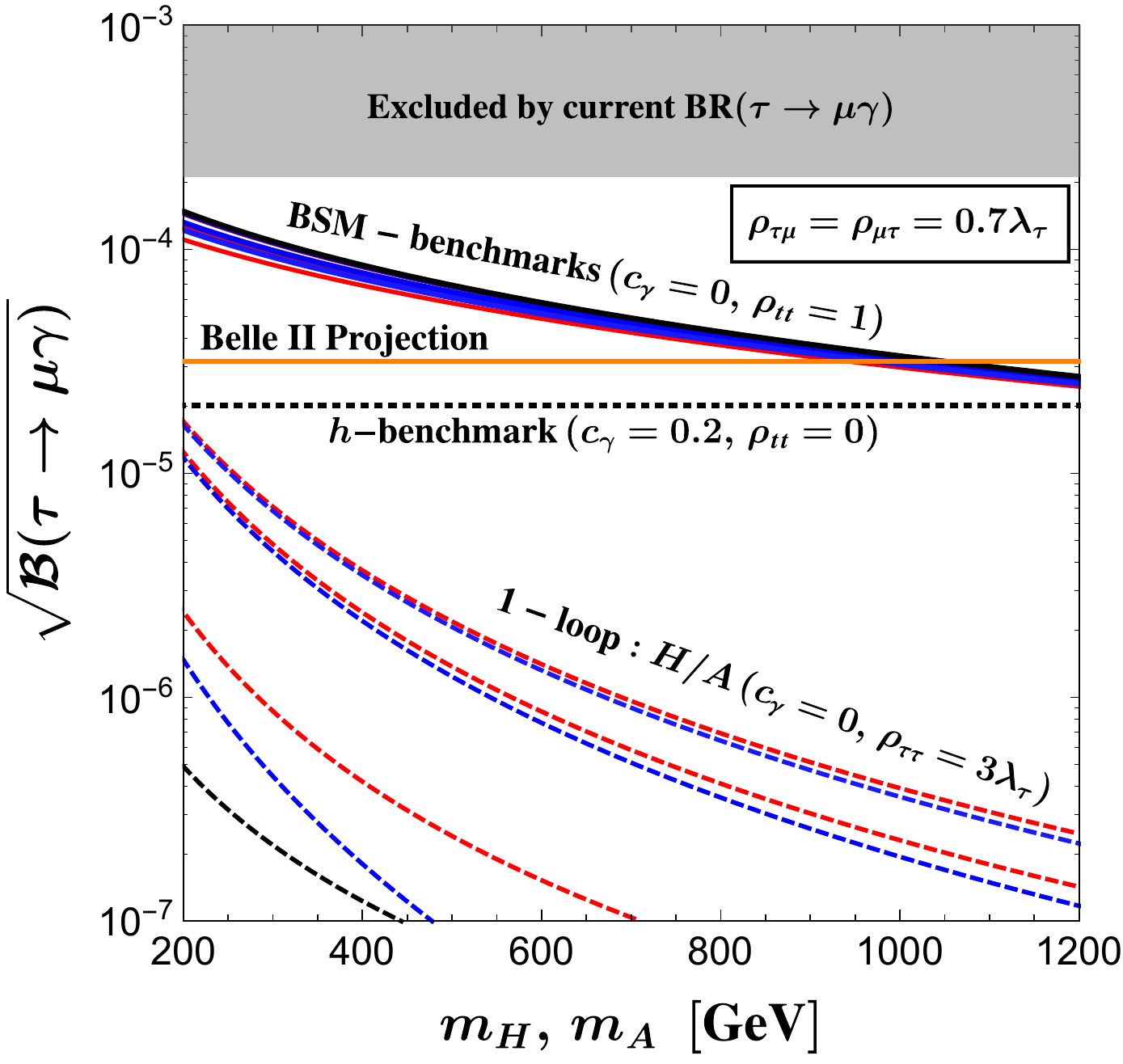}
\includegraphics[width=0.453 \textwidth]{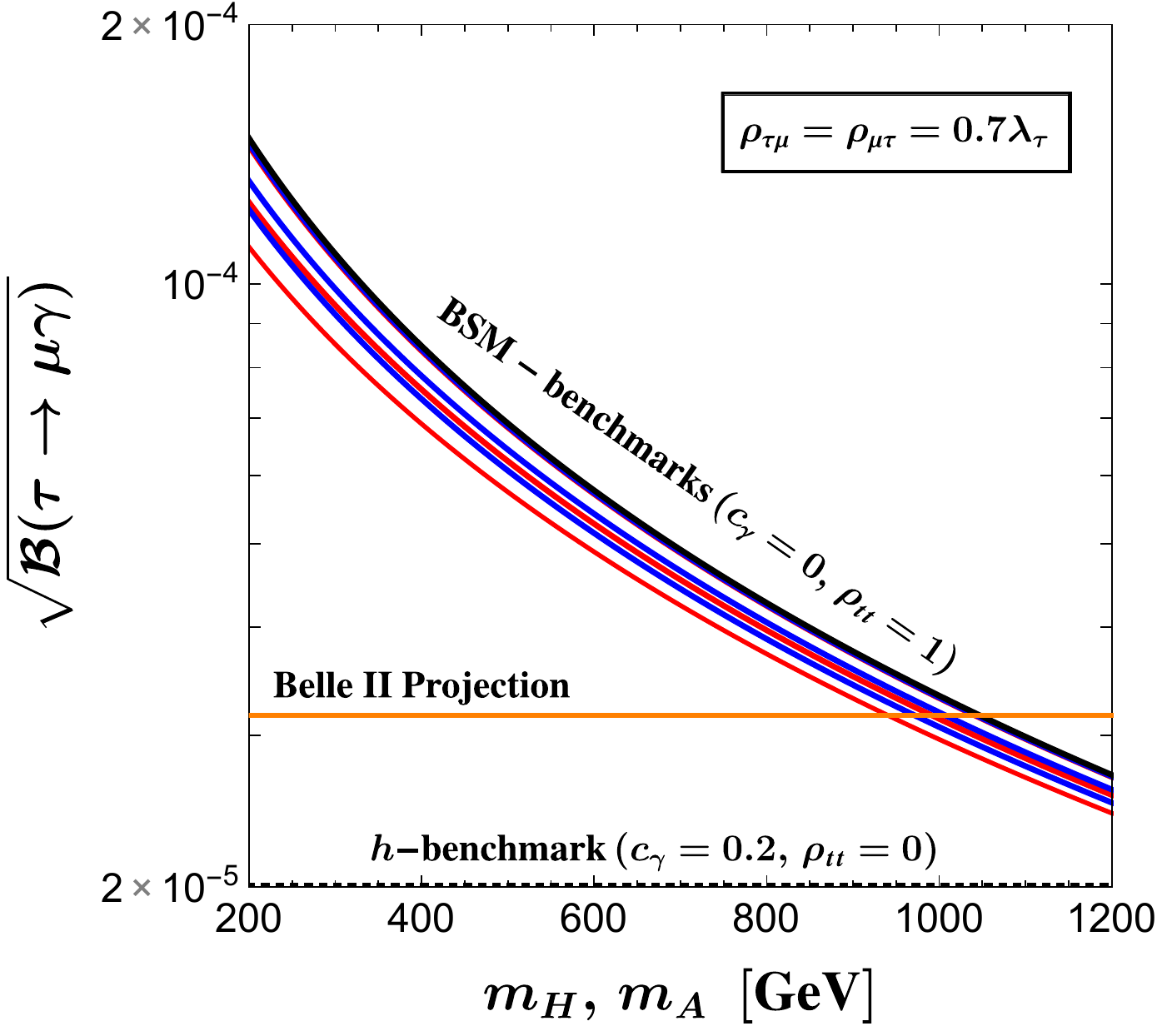}
\caption{
Comparison of the three benchmark scenarios (see text for details). 
Black curves are for the degenerate case of $m_H = m_A$.  
Red (blue) curves show variation in $m_H \,(m_A)$, with 
the other scalar mass $m_A\,(m_H)$ heavier by $ 10, 100, 200 $ GeV. }
\label{bench}
\end{figure*}

We find that the extra Yukawa couplings can always be
 normalized against the Yukawa couplings in SM, namely
\begin{align}
 \hat\rho_{3j}^f = \rho_{3j}^f/\lambda_3^f,
\label{rho-hat}
\end{align}
and perhaps Nature hints at such ``normalization''.
After all, the extra Yukawa matrix {\boldmath $\rho$}$^f$ can be viewed
as the orthogonal combination of two Yukawa matrices with respect to the mass matrix.
Along this thread, we shall take throughout this work
\begin{align}
 \rho_{32}^f, \rho_{33}^f = {\cal O}(\lambda_3^f),
\label{rho-order}
\end{align}
as our working assumption, which is the most reasonable one without tuning, 
given that $\lambda_3^f$ and {\boldmath $\rho$}$^f$ emerge from 
the procedure of diagonalizing the mass matrix.
The two Yukawa matrices should share the mass-mixing
hierarchy structure~\cite{Hou:2017hiw}.
For this reason, we illustrate with $\rho_{\tau\mu}$
and $\rho_{\tau\tau}$ values not exceeding $3\lambda_\tau$.
We remark that $\hat \rho_{\tau\tau}$ is precisely $\tan\beta_\tau$
as defined in Ref.~\cite{Davidson:2010xv}, up to a sign.
But Eq.~(\ref{rho-order}) should make clear that,
while $\hat \rho_{\tau\tau}$ and $\hat \rho_{tt}$ are both ${\cal O}(1)$,
their actual values could differ by an order of magnitude,
and should be determined by experiment.

Thus, besides scalar masses, the parameters that enter are:
$\rho_{\tau\mu}$ (overall and factorized), 
$\rho_{\tau\tau}$ (one-loop), 
$\rho_{tt}$ (two-loop),
and the $h$--$H$ mixing parameter $c_\gamma$.
Although $c_\gamma$ is expected small, its uncertain value is 
relevant in bringing in the extra Yukawa couplings of $h$ 
that can interfere with the leading two-loop top effect, as we now elucidate.
We conservatively take $c_\gamma = 0.2$ as its maximal value.

\section{(Less) Relevant Contributions}

Having clarified the natural setting of 
$\rho_{\tau\mu} = {\cal O}(\lambda_\tau)$
and $|c_\gamma| \lesssim 0.2$, we see that
$h \to \tau\mu$ search at the LHC would continue to probe this space.
Still, as there are multiple parameters that enter $\tau \to \mu\gamma$,
one needs to discern relevant from less relevant parameters and processes.
It is well known~\cite{Chang:1993kw} that, so long that 
$\rho_{tt} \sim \lambda_t \cong 1$ (Eq.~(\ref{rho-order})),
the two-loop mechanism is by far the leading effect. 
But what about the other two type of diagrams in Fig.~\ref{feyndiag}.
We propose two ``benchmarks'' to elucidate 
the leading and subleading effects, which then clarifies that,
in contrast with the much larger $\rho_{\tau\mu}, \rho_{\tau\tau}$ values taken in the past,
the one-loop diagram cannot really be probed by Belle~II
under the rule of thumb of Eq.~(\ref{rho-order}).

We define the ``BSM-benchmark'' as setting $c_\gamma = 0$ 
in the two-loop mechanism to decouple $h$, 
and take $\rho_{tt} = 1$, as larger values tend to run into 
flavor constraints~\cite{Hou:2019grj,Altunkaynak:2015twa}, 
which we shall not explore in detail here.
This benchmark captures the BSM effect from extra top Yukawa
couplings of $H, A$, and would stand alone in the alignment limit, 
 when the $\rho_{tt}$ phase no longer matters.
We plot the $\sqrt{{\cal B}(\tau \to\mu\gamma)}$ in Fig.~\ref{bench}, 
where we set $\rho_{\tau\mu} = \rho_{\mu\tau} = 0.7\lambda_\tau$
 (reason clarified below), which is conservative.
The current bound on $\tau \to\mu\gamma$ is the shaded region,
while the (conservatively) projected Belle~II limit of $10^{-9}$ is the horizontal solid line.
It is interesting that, even for the conservative value of $\rho_{\tau\mu} = 0.7\lambda_\tau$,
this BSM-benchmark can itself be readily probed by Belle~II.

Conversely, if we set $\rho_{tt}$ to zero, then the leading
two-loop effect vanishes, but the two-loop top still has 
an amplitude proportional to $\lambda_t\, c_\gamma$ coming from the $h$ boson, 
and similarly through the two-loop $W$ diagram, also with $c_\gamma$ dependence. 
Combining these $m_h$-dependent effects and calling it the ``$h$-benchmark, 
its $\sqrt{{\cal B}(\tau \to\mu\gamma)}$ is also plotted in Fig.~\ref{bench}
as the dotted line, taking the conservative maximal value of $c_\gamma = 0.2$,
which implies $\rho_{\tau\mu} = 0.7\lambda_\tau$
as maximally allowed by Eq.~(\ref{B_h-taumu}).
We see from Fig.~\ref{bench} that, if stand-alone, this $h$-benchmark is
out of Belle~II reach.
This line actually does not depend on detailed values of $c_\gamma$ or $\rho_{\tau\mu}$,
but depends only on the bound of Eq.~(\ref{B_h-taumu}), 
which follows from Eq.~(\ref{h-taumugamCMS}),  
the current bound~\cite{Sirunyan:2017xzt} on $h \to \tau\mu$.
This is because the $h$-benchmark is also proportional to $|\rho_{\tau\mu}\,c_\gamma|^2$.
Thus, the CMS bound on $h \to \tau\mu$ excludes the possibility of observing the
two-loop effect without the participation of the extra top Yukawa coupling, $\rho_{tt}$!
We enlarge this branching ratio region and display in Fig.~~\ref{bench} (right),
which can be used to understand our numerical discussion in the next Section.

We see from Fig.~\ref{bench} that, if one has relatively light
extra neutral scalars ($\lesssim 300$ GeV), then
the effect from ``BSM-benchmark'' tends to predominate.
However, as the extra scalar mass increases, say beyond 500--600 GeV,
on one hand it would require a larger fraction of full Belle~II data to probe,
on the other hand, the interference between
the BSM-benchmark and $h$-benchmark becomes important.
As the latter is real in amplitude, the phase $\varphi_{tt} = \arg \rho_{tt}$ matters, 
along with the value of $|\rho_{tt}|$, which affects the extra Higgs two-loop effect,
and the value of $c_\gamma$, which controls the effect of $h$.

\begin{figure*}[t]
\center
\includegraphics[width=0.3 \textwidth]{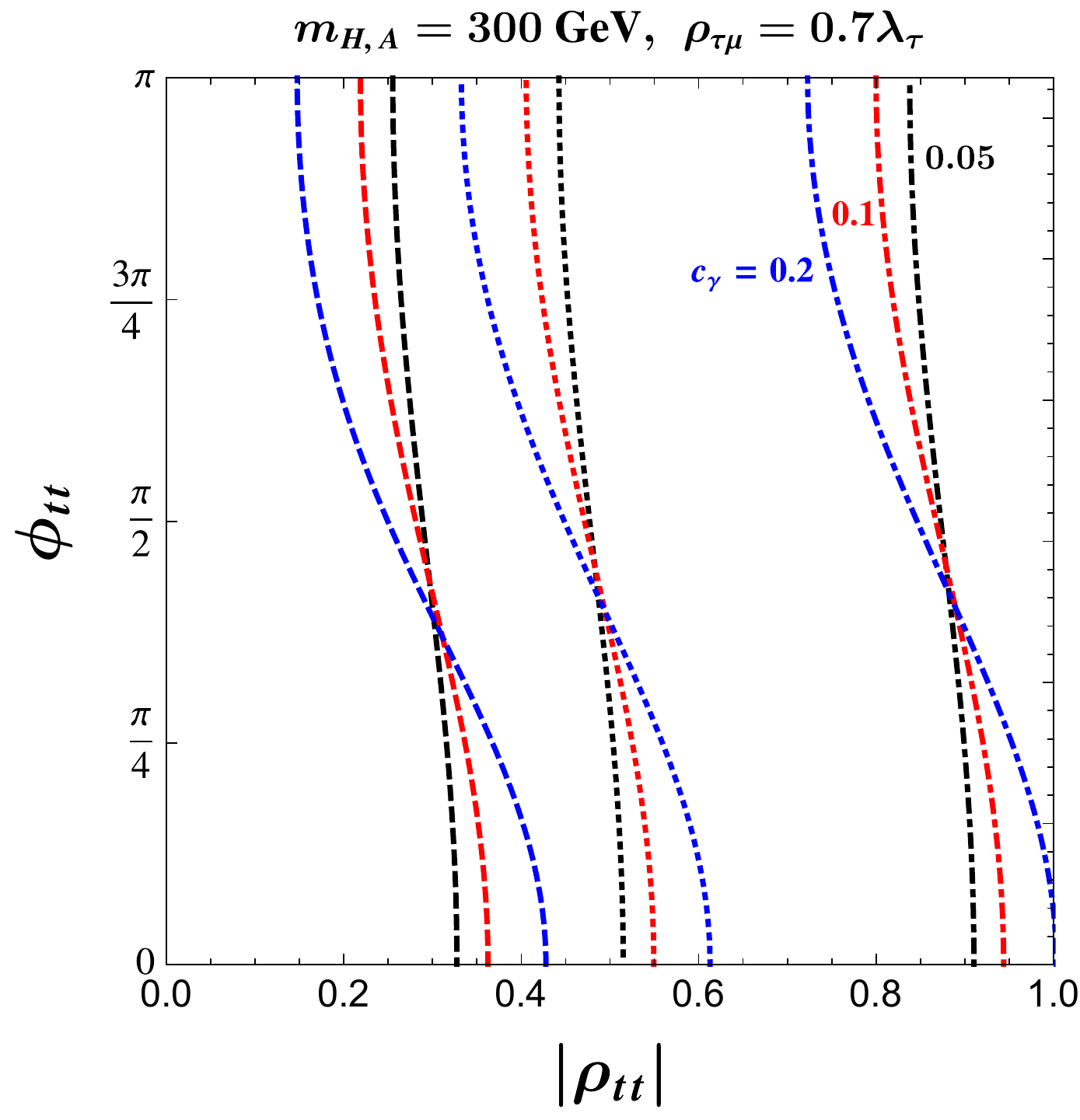}
\includegraphics[width=0.3 \textwidth]{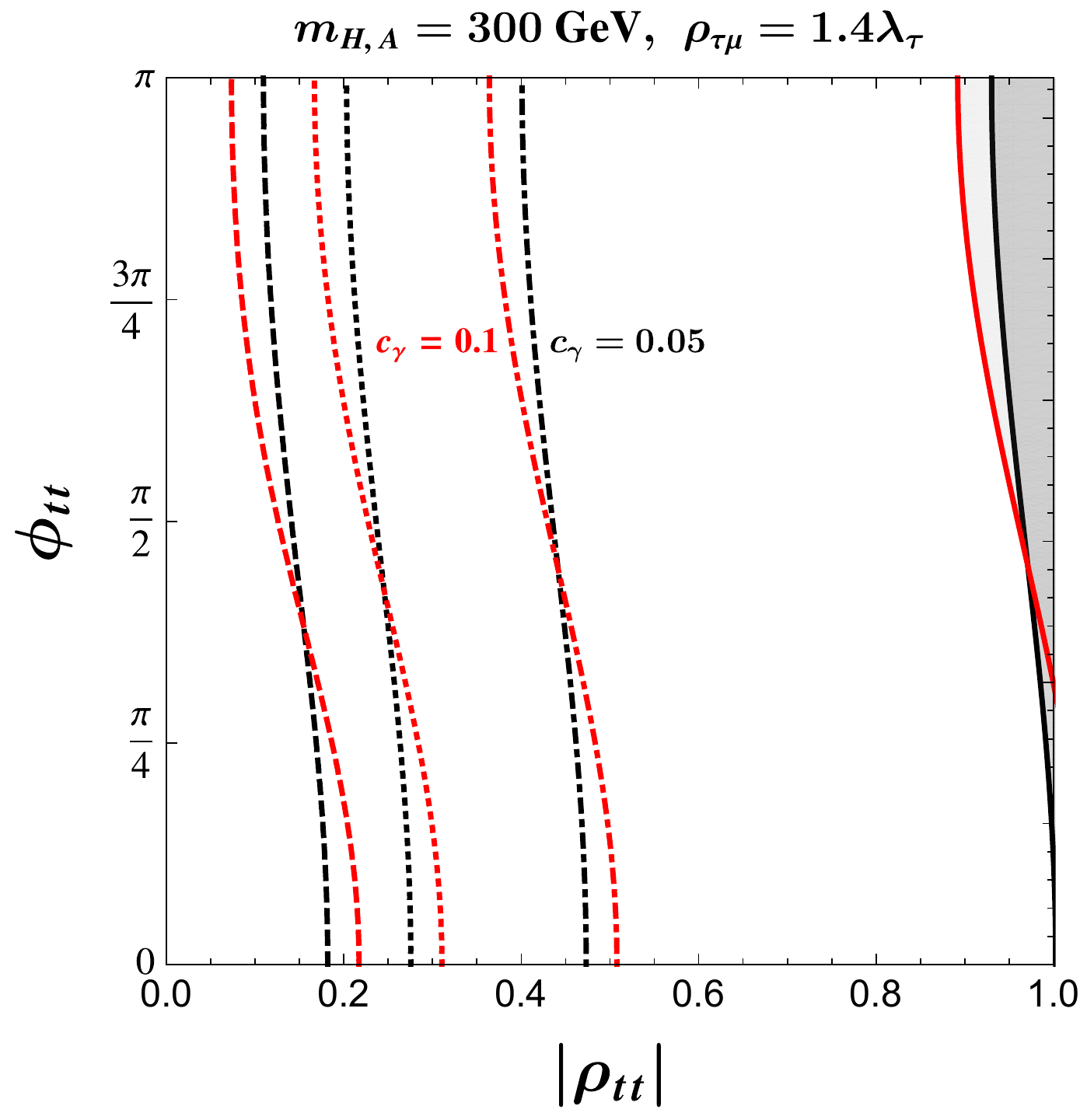}
\includegraphics[width=0.3 \textwidth]{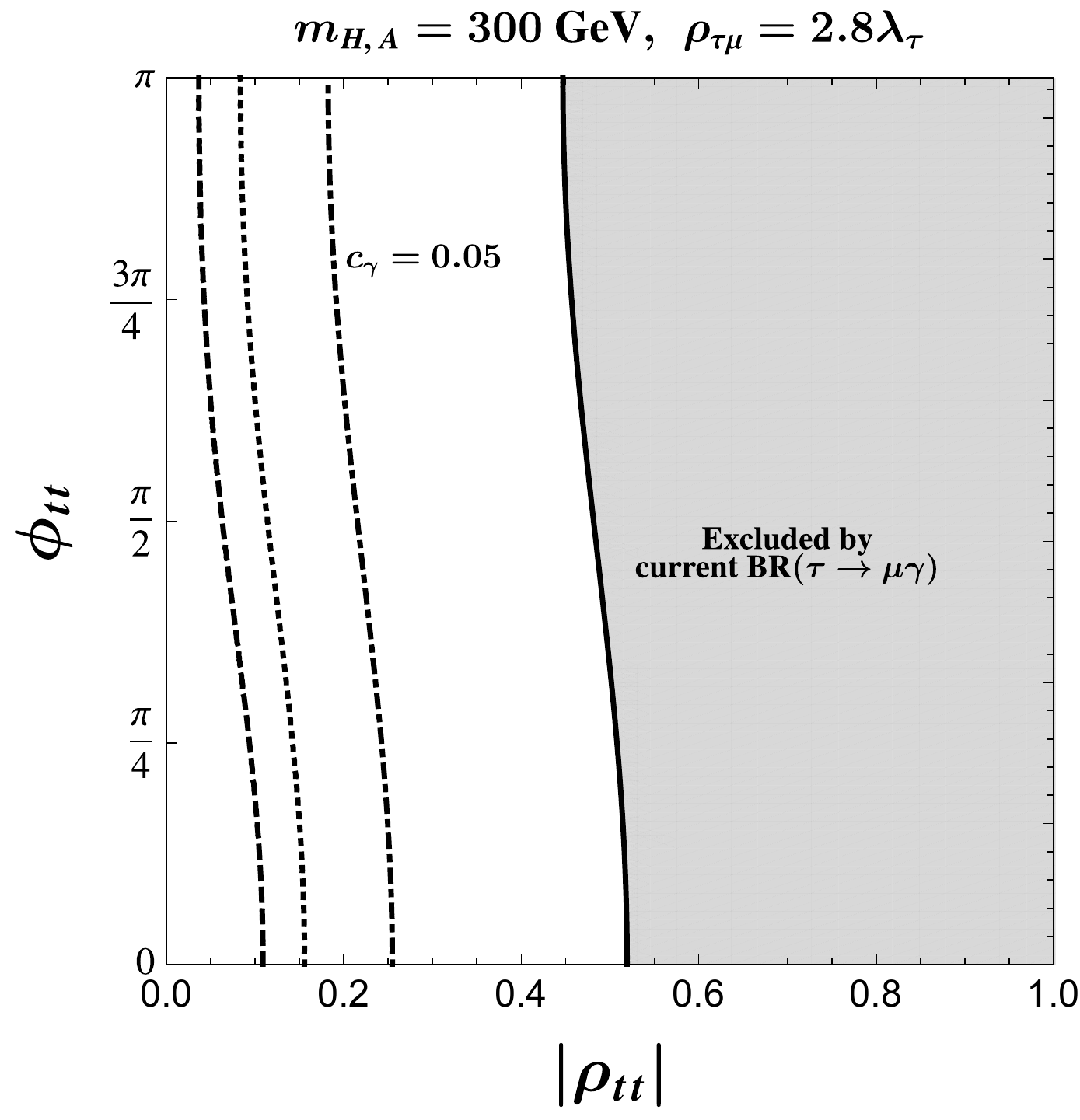}
\vskip0.1cm
\includegraphics[width=0.3 \textwidth]{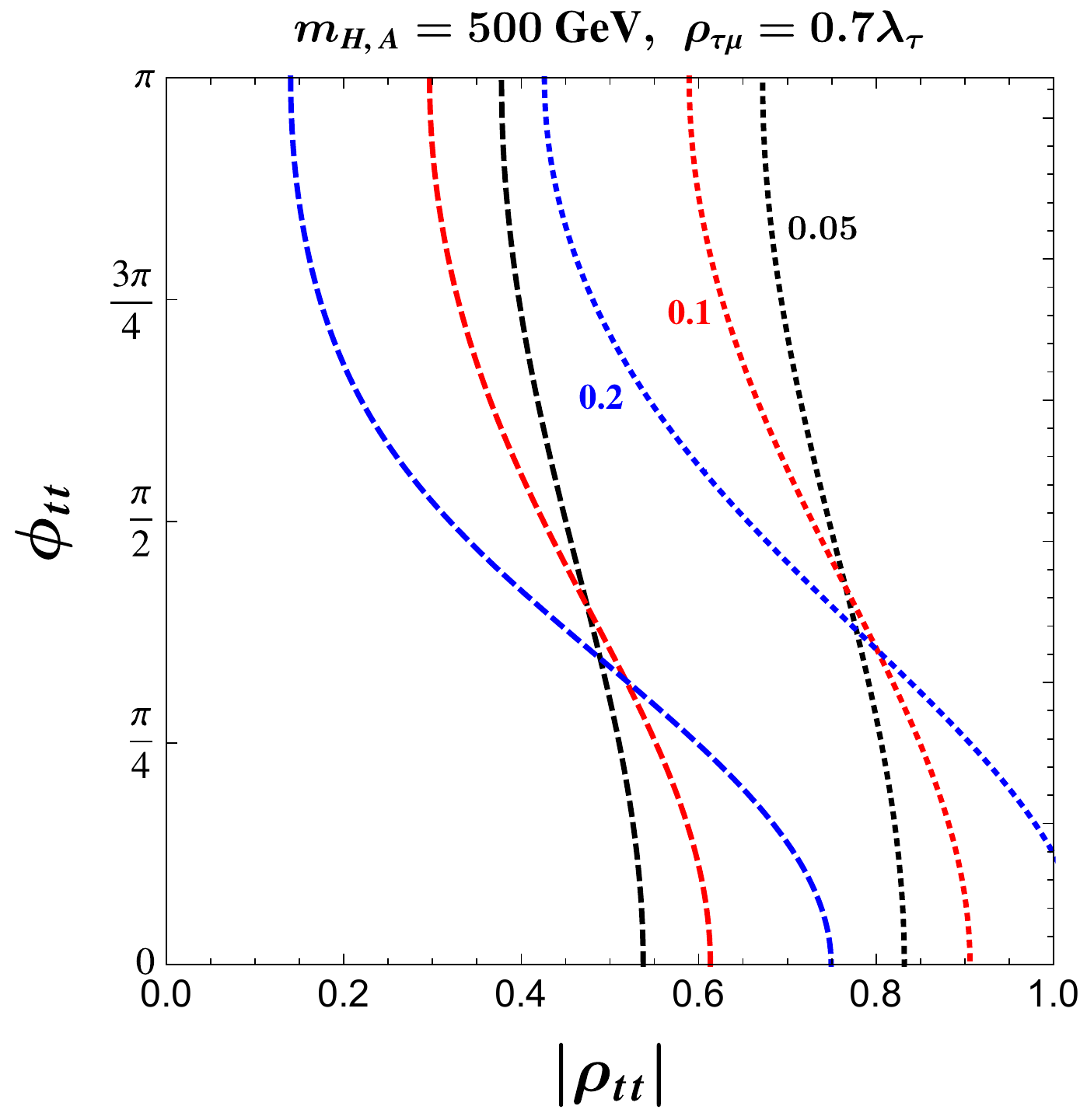}
\includegraphics[width=0.3 \textwidth]{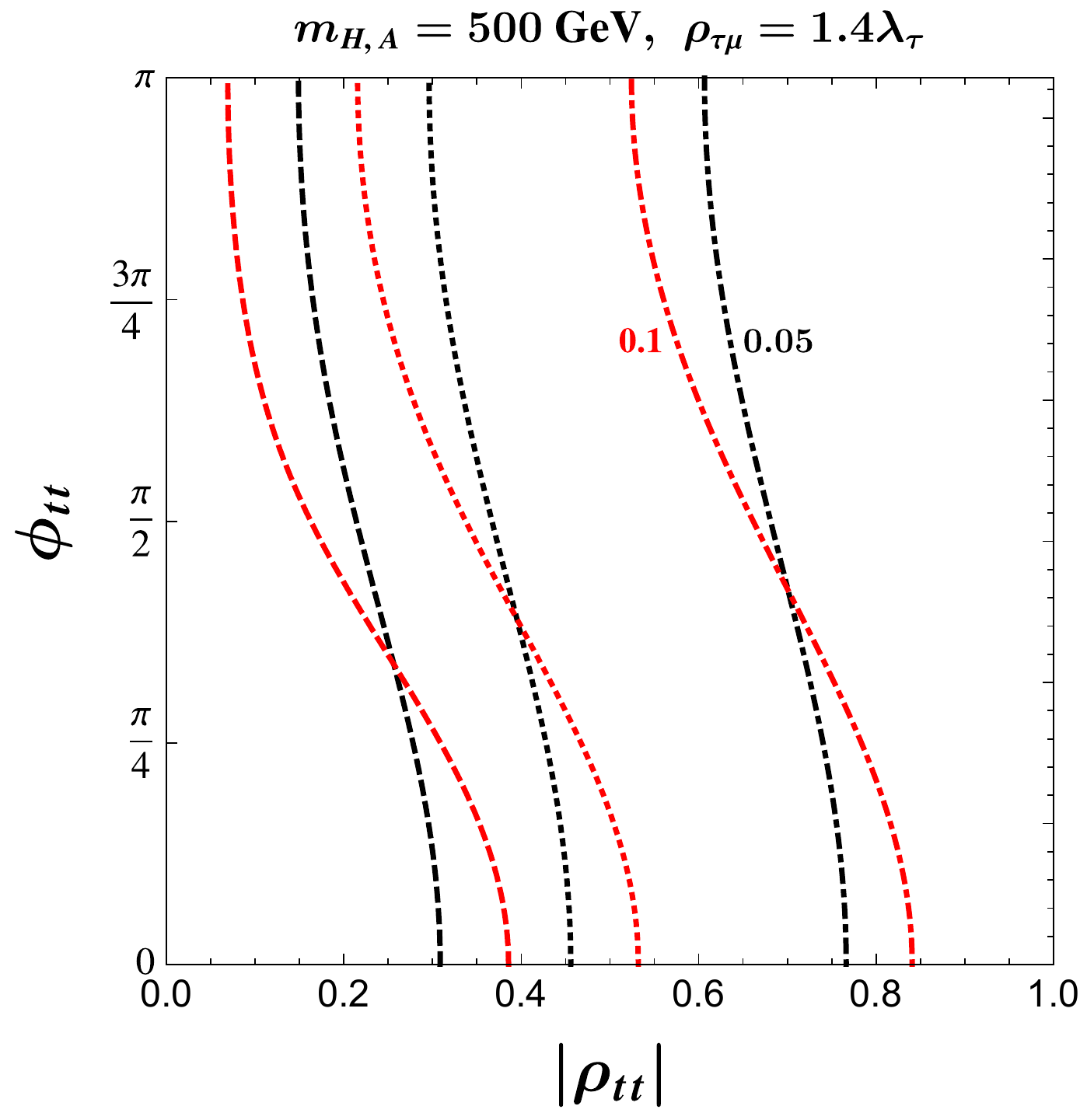}
\includegraphics[width=0.3 \textwidth]{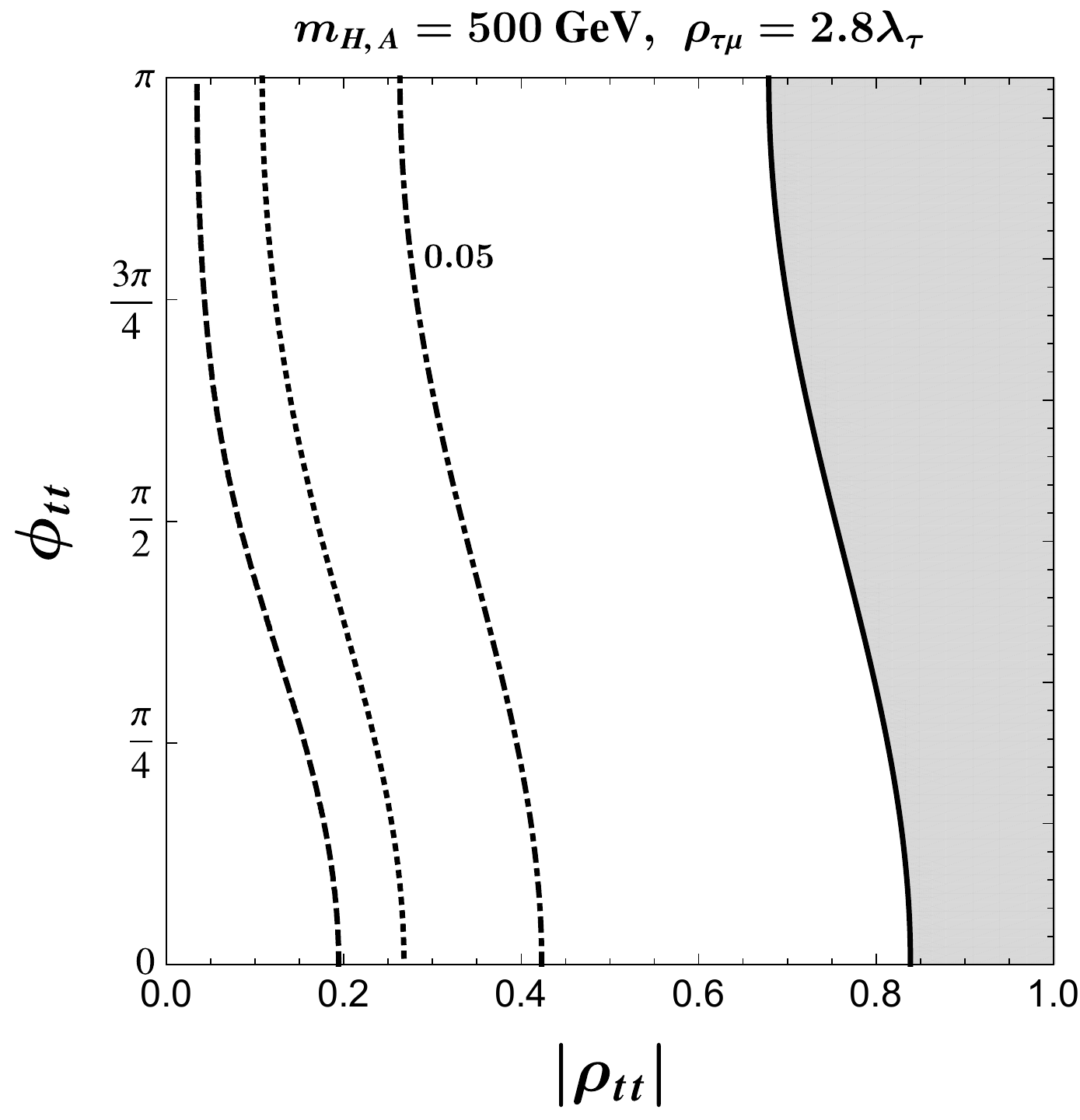}
\caption{
For $m_{H, A} = 300$ (500) GeV,
the upper (lower) plots are for the 3, 10 and 50 ab$^{-1}$ Belle~II data reach,
plotted in the $|\rho_{tt}|$--$\varphi_{tt}$ plane. 
For the lower $\rho_{\tau\mu} = 0.7\lambda_\tau$ value, 
three curves for allowed $c_\gamma$ values are illustrated,
which reduces to just one low $c_\gamma$ value for 
the larger $\rho_{\tau\mu} = 2.8\lambda_\tau$.
The shaded region is excluded by Eq.~(1).
See text for further discussion.}
\label{rhottconstraint}
\end{figure*}

Finally, we exhibit the $\sqrt{{\cal B}(\tau \to\mu\gamma)}$ of 
the one-loop $\tau$ effect in Fig.~\ref{bench},
where $\rho_{\tau\tau}$ can also carry a phase,
and we take the nominally largest value of $|\rho_{\tau\tau}| = 3\lambda_\tau$
that satisfies Eq.~(\ref{rho-order}).
It is known that the effect of $H$ and $A$  strongly cancel each other 
when degenerate (black dashed curve), 
but the cancellation weakens when degeneracy is lifted.
We give three sets of dashed curves, where red (blue) corresponds to 
$m_H$ ($m_A$) on real axis, with the other neutral scalar heavier by 10, 100, 200 GeV
(this is done also for the two-loop BSM-benchmark, where effect is minor).
For a given scalar $m_H$ ($m_A$) mass, the one-loop effect
varies by more than one order of magnitude as the splitting increases.
In general, the amplitude is far below even the $h$-benchmark,
except for rather light scalars ($\lesssim 300$~GeV).
Thus, we see that Belle~II would not have the ability to probe 
the one-loop $\tau$ contribution, that it is more than a nuisance effect.
In the next section, we neglect the one-loop effect in our illustrations,
as it just smears the projections at small $|\rho_{tt}|$,
but cannot be discerned by Belle II.

%

\section{\boldmath Interplay of  $h \to \tau\mu$ and $\tau \to \mu\gamma$}

We have exhibited in Fig.~\ref{bench} 
the BSM-benchmark, which illustrates the two-loop effect from $H$ and $A$
 with near maximal $|\rho_{tt}| = 1$, 
and the $h$-benchmark, which illustrates the two-loop effect of $h$
 with near maximal $c_\gamma = 0.2$.
The strength $|\rho_{tt}|$ --- and phase $\varphi_{tt}$ ---
and value of $c_\gamma$ (proximity to alignment limit)
together determine the strength of interference between the leading and subleading effects.
We have shown that the one-loop $\tau$ effect is less than subleading, 
which we shall ignore in the following numerical illustration.

Of course, the strength of $\rho_{\tau\mu}$ determines the overall
scale for the branching fraction, as it factorizes and one cannot probe its phase. 
Together with $c_\gamma$, $|\rho_{\tau\mu}|$ is constrained
by the bound on ${\cal B}(h \to \tau\mu)$, Eq.~(\ref{B_h-taumu}).
For instance, our near maximal value of $c_\gamma = 0.2$ for the $h$-benchmark
allows only $\rho_{\tau\mu} \lesssim 0.7\lambda_\tau$,
while $c_\gamma = 0.1, 0.05$ can allow the larger ranges of
$\rho_{\tau\mu} \lesssim 1.4\lambda_\tau, 2.8\lambda_\tau$, respectively.
For the alignment limit case of $c_\gamma = 0$, one recovers the BSM-benchmark,
which scales with $|\rho_{\tau\mu}\rho_{tt}|^2$, and can be read off from Fig.~\ref{bench}.

To illustrate the interference effect between 
the leading $H, A$ with subleading $h$ contributions and the role played by $\varphi_{tt}$,
we plot in Fig.~\ref{rhottconstraint} the future reach of Belle~II data 
at 3, 10 and 50 ab$^{-1}$ in the $|\rho_{tt}|$--$\varphi_{tt}$ plane, for the three values of 
$\rho_{\tau\mu} = 0.7\lambda_\tau, 1.4\lambda_\tau, 2.8\lambda_\tau$, respectively.
As seen from Fig.~\ref{bench}, the BSM-benchmark does not depend
strongly on $m_H$--$m_A$ splitting, so we will use 
a common $m_{H, A}$ mass value, taken as 300 and 500 GeV. 
It is illustrated e.g. in Ref.~\cite{Ghosh:2019exx} that large parameter space in 
Higgs potential is allowed by the electroweak precision $T$-parameter
and other considerations.

Let us start with the upper left plot in Fig.~\ref{rhottconstraint},
which is for the conservative value of $\rho_{\tau\mu} = 0.7\lambda_\tau$
and relatively light $m_H, m_A \simeq 300$ GeV.
From Fig.~\ref{bench} one can easily understand that the current
bound on $\tau \to \mu\gamma$, Eq.~(1), does not put a constraint
on the displayed parameter space, but can be probed as data accumulates 
at Belle~II, where the three sets of curves correspond to 3, 10 and 50 ab$^{-1}$.
Each set of curves is further illustrated with three curves 
that correspond to $c_\gamma = 0.05, 0.1, 0.2$ allowed by Eq.~(\ref{B_h-taumu}),
i.e. the bound from Eq.~(2). The curves are all of similar shape,
and the dependence on $\varphi$ illustrate the interference of $H, A$
with the $h$ effects, which is richer than the 
real value of $\rho_{tt}$ assumed in Ref.~\cite{Hou:2019grj}.
For the larger value of $\rho_{\tau\mu} = 1.4\lambda_\tau$,
$c_\gamma = 0.2$ becomes excluded, so we illustrate with 
two curves for each projected data value.
The smaller  $c_\gamma$ means the $h$ effect is reduced, hence the interference weakens,
while the current bound of Eq.~(1) starts to cut into the $|\rho_{tt}|$ parameter space
as an effect through the ``BSM-benchmark''.
For the near maximal $\rho_{\tau\mu} = 2.8\lambda_\tau$,
only the small  $c_\gamma = 0.05$ is allowed, hence we show 
only one curve for each data value in the right figure,
and the current bound of Eq.~(1) now cuts deeper into $|\rho_{tt}|$ parameter space.

The lower plots of Fig.~\ref{rhottconstraint} are for heavier $m_H, m_A = 500$ GeV,
hence the contribution from the ``BSM-benchmark" is weakened, resulting in 
stronger interference due to the relative importance of the ``$h$-benchmark'' contribution.
For $\rho_{\tau\mu}$ at the conservative $0.7\lambda_\tau$, 
$\tau\to \mu\gamma$ does not yet start to probe the $|\rho_{tt}|$ parameter space 
even with 3~ab$^{-1}$.
For $\rho_{\tau\mu} = 1.4\lambda_\tau$, the current bound
is still ineffective, but 3 ab$^{-1}$ would cut into $|\rho_{tt}|$ parameter space,
while for the relatively large $\rho_{\tau\mu} = 2.8\lambda_\tau$, 
even the current bound excludes some $|\rho_{tt}|$ parameter space.

Our figures project the discovery potential of $\tau \to \mu\gamma$ by Belle II, 
as constrained by $h \to \tau\mu$ under our working assumption of 
$\rho_{\tau\mu} = {\cal O}(\lambda_\tau) \sim 0.01$.
The parameter space is substantial, so long that
$\rho_{\tau\mu}/\lambda_\tau$ is not far below 1,
and the extra Higgs mass scale does not approach decoupling.

\section{Discussion and Summary}

The constraint of Eq.~(\ref{B_h-taumu}), which arises from $h \to \tau\mu$ search at the LHC,
should improve in the next couple of years when the full Run~2 data is analyzed.
It would likely drop further, which would imply that our 
``$h$-benchmark'' line in Fig.~2 would drop.
This would mean the interference effect as exhibited in Fig.~3 would shrink further, 
and one has less access to the phase $\varphi_{tt}$.
However, it is not impossible that a hint emerges for $h \to \tau\mu$,
which would suggest that neither $\rho_{\tau\mu}$ nor $c_\gamma$ vanish,
and would heighten the interest in $\tau \to \mu\gamma$ search at Belle~II.
Assuming no hint for signal,  combining the full Run 2+3 dataset of ATLAS and CMS
and scaling naively by statistics, one can probe down to $0.05\%$, 
compared with 0.25\% in Eq.~(\ref{B_h-taumu}).
One would then be close to the ``BSM-benchmark'' scenario.
If we happen to be rather close to the alignment limit,
then the constraint on $\rho_{\tau\mu}$ is alleviated, 
with $\tau \to \mu\gamma$ probing $|\rho_{\tau\mu}\rho_{tt}|^2$,
and Belle~II would still have wide discovery potential.

It should be noted that exotic Higgs bosons as light as 300 GeV is 
{\it not} ruled out~\cite{Hou:2018zmg}. 
There is in fact a mild hint for a pseudoscalar $A$ around 400 GeV~\cite{Sirunyan:2019wph}, 
interfering with the $gg \to t\bar t$ QCD background.
It could be the $gg$ fusion production and decay of $A$ via $\rho_{tt}$,
or even $H$ that is produced via 
a purely imaginary $\rho_{tt}$ coupling~\cite{Hou:2019gpn}.
The exotic Higgs spectrum for g2HDM is largely unknown,
but 300 to 600~GeV is a preferred target zone, 
if~\cite{Hou:2017hiw} the inertial mass scale of 
the second (non-mass-giving) doublet is not far above the weak scale, 
which would be the tuned case of {\it decoupling}.
Besides $gg \to H, A \to t\bar t$~\cite{Hou:2019gpn}, $t\bar c$~\cite{Altunkaynak:2015twa},
proposed searches such as $cg \to t\,H/A \to tt\bar c$, $tt\bar t$~\cite{Kohda:2017fkn}
and the recently proposed $cg \to bH^+$~\cite{Ghosh:2019exx} process,  
give rise to signatures of same-sign top with jets, triple-top, 
and single top with two $b$-jets. 
Especially if the mass scale is below 400 GeV, we should have
good hope of learning the mass spectrum in the coming years.
Note that the three signatures above all require sizable $\rho_{tc}$ for production, 
which is in line with our working assumption of Eq.~(\ref{rho-order}).
Furthermore, $\rho_{tc}$ at ${\cal O}(1)$ can {\it also drive} BAU~\cite{Fuyuto:2017ewj}.
Thus, the program is well motivated.

We have illustrated that the discovery potential at Belle~II does not actually depend on 
whether a hint for $h\to \tau\mu$ emerges at the LHC, 
which is in part regulated by the strength of $c_\gamma$.
The actual value of $c_\gamma$, however, may be hard to extract.
Although ATLAS and CMS have fitted for $\cos(\beta-\alpha)$ 
in the context of 2HDM~II~\cite{Khachatryan:2016vau, Aad:2019mbh}, 
with many more parameters in g2HDM, such a fit may not be feasible
until we know more about some parameters related to the Extra Higgs,
such as mass spectrum.
We have conservatively taken the maximal value of 0.2 for $c_\gamma$,
but we do not view $c_\gamma = 0.3$ as ruled out in g2HDM.

Processes that do not depend on $c_\gamma$, such as 
electroweak baryogenesis (EWBG), i.e. generating BAU, are therefore of interest.
Back on Earth, we note that $H^+$ and $A$ couplings do not depend on $c_\gamma$.
Thus, $B^- \to \mu^- \bar\nu$ where the flavor of $\bar\nu$ is not detected,
probes the product of $\rho_{tu}\rho_{\tau\mu}$~\cite{Hou:2019uxa}. 
Although we do not advocate that $\rho_{tu}$ should also satisfy
some relation similar to Eq.~(\ref{rho-order}), we have rather poor knowledge of its value.
Ref.~\cite{Hou:2019uxa} suggests that the ratio of
${\cal B}(B \to \mu\bar\nu)/{\cal B}(B \to \tau\bar\nu)$ in g2HDM
may deviate from the SM expectation of 0.0045, a value that is shared by 2HDM~II. 
If such a result is found, which could emerge relatively early with Belle~II, 
it would imply nonvanishing $\rho_{\tau\mu}$,
hence would also heighten the interest in $\tau\to \mu\gamma$
(as well as pursuit of the $tuh$ coupling).
One could also probe $\rho_{\tau\mu}$ via searching for
heavy $H, A \to \tau\mu$~\cite{Hou:2019grj}.
While such search is clearly worthy~\cite{Sirunyan:2019shc}, 
it runs again branching ratio suppression due to the
likely dominance of $t\bar t$ and $t\bar c$ decay modes in g2HDM.
Although we do not think that Belle~II could effectively probe
$\rho_{\tau\tau}$ through the one-loop $\tau \to \mu\gamma$ effect,
$\rho_{\tau\tau}$ can be probed at the LHC in principle,
both via deviations from SM rate for $h \to \tau\tau$ by $h$-$H$ mixing,
or by search for heavy $H, A \to \tau\tau$~\cite{Aaboud:2017sjh, Sirunyan:2018zut},
but it might not be better than the $\tau\mu$ final state.

In summary, we analyze the outlook for $\tau$LFV search via 
the $h \to \tau\mu$ and $\tau \to \mu\gamma$ processes,
which appears quite promising in the general 2HDM.
The $h \to \tau\mu$ process probes the product $\rho_{\tau\mu}c_\gamma$,
where $\rho_{\tau\mu}$  is the extra flavor changing neutral Higgs coupling, 
and $c_\gamma$ is the $CP$-even Higgs mixing angle,
which is expected to be small by the phenomenon of alignment. 
But whether or not a hint emerges with Run 2+3 data,
our working assumption that $\rho_{\tau\mu} = {\cal O}(\lambda_\tau)$
and $\rho_{tt} = {\cal O}(\lambda_t)$ makes $\tau \to \mu\gamma$
very interesting at Belle~II, with broad parameter range for discovery.
If Nature provides a finite $c_\gamma$ that is on 
the larger side, on one hand it increases the likelihood that 
$h \to \tau\mu$ may emerge, on the other hand, 
the interference of $h$ with $H, A$ effects in $\tau\to \mu\gamma$ decay
in principle probes the phase of $\rho_{tt}$.
We look forward to the unfolding of these two search modes
in the coming decade.

\vskip0.2cm
\noindent{\bf Acknowledgments} \
We thank K.-F. Chen, S. Davidson, M. Kohda and M. Nakao for discussions.
This research is supported by 
MOST 106-2112-M-002-015-MY3, 108-2811-M-002-626, 
and NTU 109L104019.



\begin{thebibliography}{99}



\bibitem{PDG}
  M. Tanabashi {\it et al.} [Particle Data Group], Phys.\ Rev.\ D {\bf 98}, 030001 (2018).
%
\bibitem{Aubert:2009ag} 
  B.~Aubert {\it et al.} [BaBar Collaboration],
  Phys.\ Rev.\ Lett.\  {\bf 104}, 021802 (2010).
%
\bibitem{Hayasaka:2007vc} 
  K.~Hayasaka {\it et al.} [Belle Collaboration],
  Phys.\ Lett.\ B {\bf 666}, 16 (2008).
%
\bibitem{Kou:2018nap} 
  E.~Kou, P.~Urquijo {\it et al.} [Belle-II Collaboration],
  PTEP {\bf 2019}, 123C01 (2019).


 \bibitem{h125_discovery}
%
  G.~Aad {\it et al.}  [ATLAS Collaboration],
  Phys.\ Lett.\ B {\bf 716}, 1 (2012);~
%
  S.~Chatrchyan {\it et al.}  [CMS Collaboration],
  {\it ibid.} 
  \ B {\bf 716}, 30 (2012).
%
\bibitem{Khachatryan:2015kon} 
  V.~Khachatryan {\it et al.} [CMS Collaboration],
  Phys.\ Lett.\ B {\bf 749}, 337 (2015).
%
\bibitem{Sirunyan:2017xzt} 
  A.M.~Sirunyan {\it et al.} [CMS Collaboration],
  JHEP {\bf 1806}, 001 (2018).
%
\bibitem{Aad:2019ugc} 
  G.~Aad {\it et al.} [ATLAS Collaboration],
  Phys.\ Lett.\ B {\bf 800}, 135069 (2020).


%
\bibitem{Branco:2011iw} 
  G.C.~Branco,  P.~M.~Ferreira, L.~Lavoura, M.~N.~Rebelo, M.~Sher and J.~P.~Silva,
  Phys.\ Rept.\  {\bf 516}, 1 (2012).
%
\bibitem{Hou:1991un} 
  W.-S.~Hou,
  Phys.\ Lett.\ B {\bf 296}, 179 (1992).
%
\bibitem{Vicente:2019ykr} 
  A.~Vicente,
  Front.\ in Phys.\  {\bf 7}, 174 (2019).
%
\bibitem{Chang:1993kw} 
  D.~Chang, W.-S.~Hou and W.-Y.~Keung,
  Phys.\ Rev.\ D {\bf 48}, 217 (1993).
%
\bibitem{Han:2000jz} 
  T.~Han and D.~Marfatia,
  Phys.\ Rev.\ Lett.\  {\bf 86}, 1442 (2001).
%
\bibitem{Davidson:2010xv} 
  S.~Davidson and G.J.~Grenier,
  Phys.\ Rev.\ D {\bf 81}, 095016 (2010).
%
\bibitem{Davidson:2005cw} 
  S.~Davidson and H.E.~Haber,
  Phys.\ Rev.\ D {\bf 72}, 035004 (2005).
%
\bibitem{Sierra:2014nqa} 
  D.~Aristizabal Sierra and A.~Vicente,
  Phys.\ Rev.\ D {\bf 90}, 115004 (2014).
%
\bibitem{Harnik:2012pb} 
  R.~Harnik, J.~Kopp and J.~Zupan,
  JHEP {\bf 1303}, 026 (2013).
%
  See also S.~Davidson and P.~Verdier,
  Phys.\ Rev.\ D {\bf 86}, 111701 (2012).
%
\bibitem{Cheng:1987rs} 
  T.-P.~Cheng and M.~Sher,
  Phys.\ Rev.\ D {\bf 35}, 3484 (1987).
%
\bibitem{Omura:2015xcg} 
  Y.~Omura, E.~Senaha and K.~Tobe,
  Phys.\ Rev.\ D {\bf 94}, 055019 (2016).
%
\bibitem{Hou:2019grj} 
  W.-S.~Hou, R.~Jain, C.~Kao, M.~Kohda, B.~McCoy and A.~Soni,
  Phys.\ Lett.\ B {\bf 795}, 371 (2019).
%
\bibitem{Sirunyan:2019shc} 
  A.M.~Sirunyan {\it et al.} [CMS Collaboration],
  arXiv:1911.10267 [hep-ex].
%
\bibitem{Fuyuto:2017ewj} 
  K.~Fuyuto, W.-S.~Hou and E.~Senaha,
  Phys.\ Lett.\ B {\bf 776}, 402 (2018).


  
%
\bibitem{Hou:2017hiw} 
  W.-S. Hou, M.~Kikuchi,
  EPL {\bf 123}, 11001 (2018).
%
\bibitem{Khachatryan:2016vau} 
  G.~Aad {\it et al.} [ATLAS and CMS Collaborations],
  JHEP {\bf 1608}, 045 (2016).
%
\bibitem{Chen:2013qta} 
  K.-F.~Chen, W.-S.~Hou, C.~Kao and M.~Kohda,
  Phys.\ Lett.\ B {\bf 725}, 378 (2013).


%
\bibitem{Altunkaynak:2015twa} 
  B.~Altunkaynak, W.-S.~Hou, C.~Kao, M.~Kohda and B.~McCoy,
  Phys.\ Lett.\ B {\bf 751}, 135 (2015).


%
\bibitem{Ghosh:2019exx} 
  D.~K.~Ghosh, W.~S.~Hou and T.~Modak,
  arXiv:1912.10613 [hep-ph].
 

%
\bibitem{Hou:2018zmg} 
  See, for example W.-S.~Hou, M.~Kohda and T.~Modak,
  Phys.\ Lett.\ B {\bf 786}, 212 (2018),
  as well as consult the PDG listings.
%
\bibitem{Sirunyan:2019wph} 
  A.M.~Sirunyan {\it et al.} [CMS Collaboration],
  arXiv:1908.01115 [hep-ex].
%
\bibitem{Hou:2019gpn} 
  W.-S.~Hou, M.~Kohda and T.~Modak,
  Phys.\ Lett.\ B {\bf 798}, 134953 (2019).
 %
\bibitem{Kohda:2017fkn} 
  M.~Kohda, T.~Modak and W.-S.~Hou,
  Phys.\ Lett.\ B {\bf 776}, 379 (2018).
%
\bibitem{Aad:2019mbh} 
  For the latest large dataset results, see G.~Aad {\it et al.} [ATLAS Collaboration],
  Phys.\ Rev.\ D {\bf 101}, 012002 (2020); and references therein.
%
\bibitem{Hou:2019uxa} 
  W.-S.~Hou, M.~Kohda, T.~Modak and G.-G.~Wong,
  Phys.\ Lett.\ B {\bf 800}, 135105 (2020).
%
\bibitem{Aaboud:2017sjh} 
  M.~Aaboud {\it et al.} [ATLAS Collaboration],
  JHEP {\bf 1801}, 055 (2018).
%
\bibitem{Sirunyan:2018zut} 
  A.M.~Sirunyan {\it et al.} [CMS Collaboration],
  JHEP {\bf 1809}, 007 (2018).

\end{thebibliography}
\end{document}